\documentclass[a4paper]{jpconf}
\usepackage{graphicx}
\begin{document}
\title{On the formation of molecules and solid-state compounds from the AGB to
the PN phases}

\author{D. A. Garc\'{\i}a-Hern\'andez$^{1,2}$ and A. Manchado$^{1,2,3}$}
%The style for the names is initials then surname, with a comma after all but the last two names, which are separated by `and'. Initials should {\it not} have full stops. First names may be used if desired.

\address{$^1$ Instituto de Astrof\'{\i}sica de Canarias, C/ Via L\'actea s/n, E-38205 La Laguna, Spain\\
$^2$ Departamento de Astrof\'{\i}sica, Universidad de La Laguna (ULL), E-38206 La Laguna, Spain\\
$^3$ Consejo Superior de Investigaciones Cient\'{\i}ficas, Spain}

\ead{agarcia@iac.es; amt@iac.es}

\begin{abstract}
During the asymptoyic giant branch (AGB) phase, different elements are
dredge-up to the stellar surface depending on progenitor mass and metallicity.
When the mass loss increases at the end of the AGB, a circumstellar dust shell
is formed, where different (C-rich or O-rich) molecules and solid-state
compounds are formed. These are further processed in the transition phase
between AGB stars and planetary nebulae (PNe) to create more complex organic
molecules and inorganic solid-state compounds (e.g., polycyclic aromatic
hydrocarbons, fullerenes, and graphene precursors in C-rich environments and
oxides and crystalline silicates in O-rich ones). We present an observational
review of the different molecules and solid-state materials that are formed from
the AGB to the PN phases. We focus on the formation routes of complex fullerene
(and fullerene-based) molecules as well as on the level of dust processing
depending on metallicity.
\end{abstract}

\section{Introduction}

Low- and intermediate-mass (1 $<$ M $<$ 8 M$_{\odot}$) stars suffer strong mass
loss and a series of thermal pulses at the tip of the asymptotic giant branch
(AGB) phase. The interstellar medium (ISM) is thus efficiently enriched with
specific elements/isotopes and dust grains expelled by AGB stars. The principal
nucleosynthetic processes occur on the thermal pulsing (TP) AGB phase, while the
main molecular/dust processing takes place during the short transition phase
between the AGB phase and the planetary nebula (PN) stage. Carbon ($^{12}$C) and
s-process neutron-rich elements (e.g., Rb, Zr, Sr, Nd, Ba, Tc) are synthesized
and transported to the stellar surface during the TP-AGB phase. Solar
metallicity low-mass (M$<$1.5 M$_{\odot}$) AGB stars remain O-rich and they
probably do not form an observable PN, while more massive (1.5$<$M$<$4
M$_{\odot}$) AGBs are converted to C-rich (C/O$>$1) stars. The most massive
(M$>$4 M$_{\odot}$) AGBs remain also as O-rich stars because of the activation
of the hot bottom burning (HBB) process. Also, the dominant neutron source
($^{13}$C vs. $^{22}$Ne) at the s-process site depends on stellar mass, which
translates into different s-process abundances patterns (e.g., the [Rb/Zr]
ratio; see e.g., [1]). This basic AGB picture is strongly dependent on
metallicity  (e.g., [1,2,3] and references therein); e.g., the minimum
progenitor mass to efficiently activate the HBB process decreases at lower
metallicities. The higher mass AGB stars thus produce different
elements/isotopes than the lower mass AGB ones, this being imprinted in the gas
and circumstellar dust chemistry. At the end of the AGB, the more massive C-rich
and O-rich stars are also heavily obscured by their thick circumstellar dust
envelopes (CSE), experiencing a phase of total obscuration in their way to form
PNe; they are thus only accesible in the infrared (IR) and millimeter spectral
ranges.

AGB and post-AGB stars are also amazing molecular factories; i.e., more than 70
molecules have been detected in AGB CSE through their rotational transitions
from the IR to millimeter wavelengths (e.g., [4]). These are mainly gas-phase
molecules: i), inorganics like SiS and AlCl; ii) organics such as H$_{2}$CO and
CH$_{3}$CN; iii) radicals (e.g., HCO$^{+}$); iv) rings (e.g., C$_{3}$H$_{2}$);
and v) chains (e.g., HC$_{9}$N). We note that the formation of most (although
not all) of these molecules may be explained by gas-phase reactions but
solid-state chemistry has to be considered also in the models; i.e., molecules
(e.g., H$_{2}$) may form on dust grains.

The CSE dust composition of sources evolving from the AGB to the PN stage
depends on the dominant star's chemistry (i.e., the C/O ratio). The C-rich
(C/O$>$1) AGB stars show SiC and amorphous carbon, although other complex and
disordered organic solids with mixed aromatic/aliphatic structures like kerogen
and coal may provide also the strong dust continuum emission (e.g., [5]). The
O-rich (C/O$<$1) AGBs, however, display amorphous silicates, weak crystalline
silicates such as piroxenes, and refractory oxides (e.g., corundum and spinel)
(e.g., [6]). On the other hand, young and evolved O-rich PNe mainly display
strong crystalline silicate features, while the aromatic infrared bands
(AIBs, generally attributed to polycyclic aromatic hydrocarbons; PAHs) and
unidentified IR emission features (UIRs) are observed in the C-rich ones.
Finally, double-chemistry PNe are characterized by both C-rich (e.g., PAHs) and
O-rich (e.g., crystalline silicates) dust features (e.g., [7]).

Several dust types can be differentiated depending on the solid-state features
observed in the {\it Spitzer Space Telescope} spectrocopic observations of
Galactic PNe (see Figure 1, left panel), which otherwise are generally
characterized by a strong IR dust continuum emission [8,9]. Carbon chemistry
(CC) and oxygen chemistry (OC) PNe display respectively C-rich and O-rich dust
features that may be aliphatic/amorphous and/or aromatic/crystalline.
Double-chemistry (DC) PNe simultaneously show C-rich (mainly aromatic) and
O-rich dust features (crystalline and/or amorphous). Finally, featureless (F)
PNe show only nebular emission lines in their IR spectra with very weak dust
continuum emission. 

\begin{figure}
\begin{center}
\includegraphics[width=2.8in]{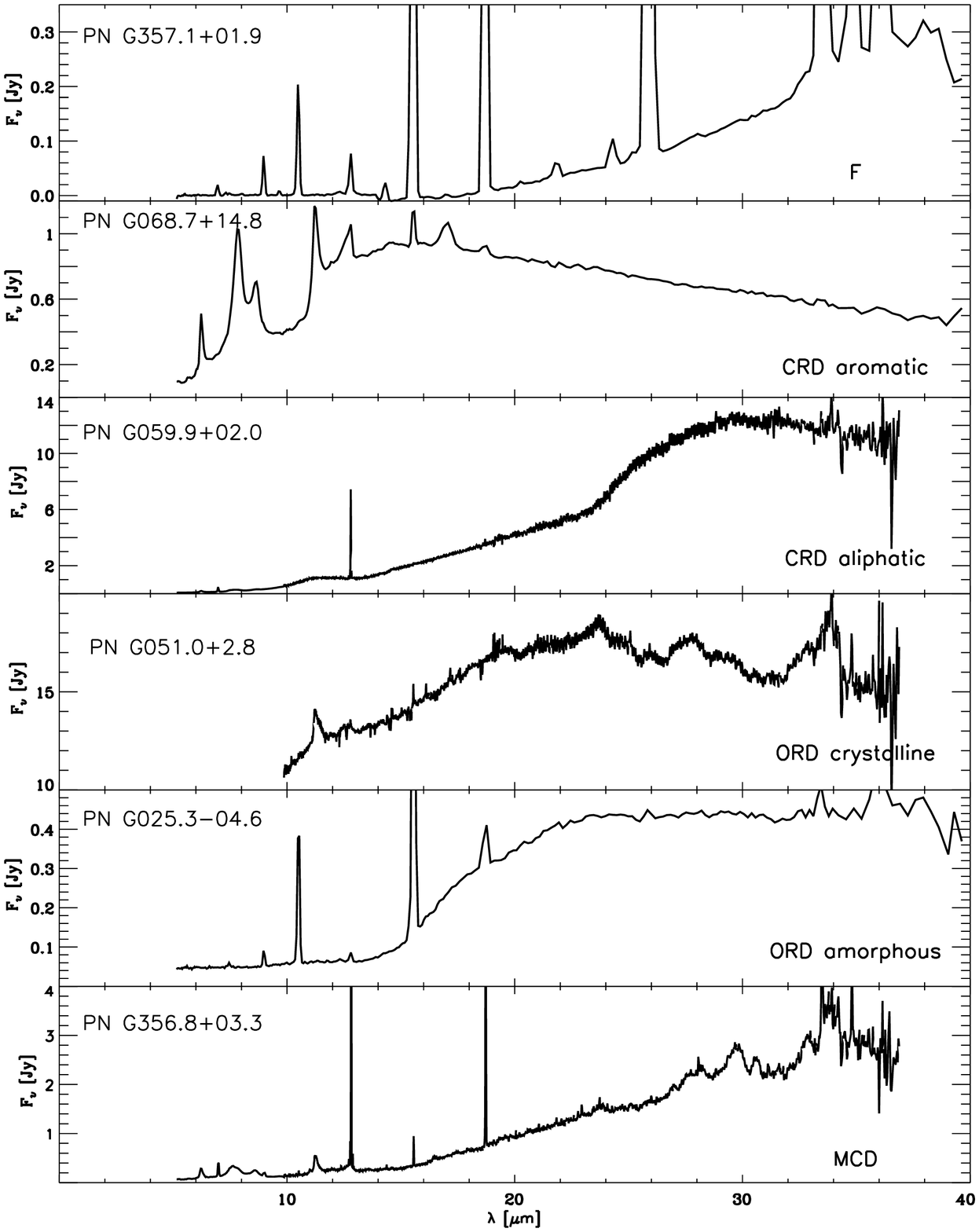}%
\includegraphics[width=3.2in]{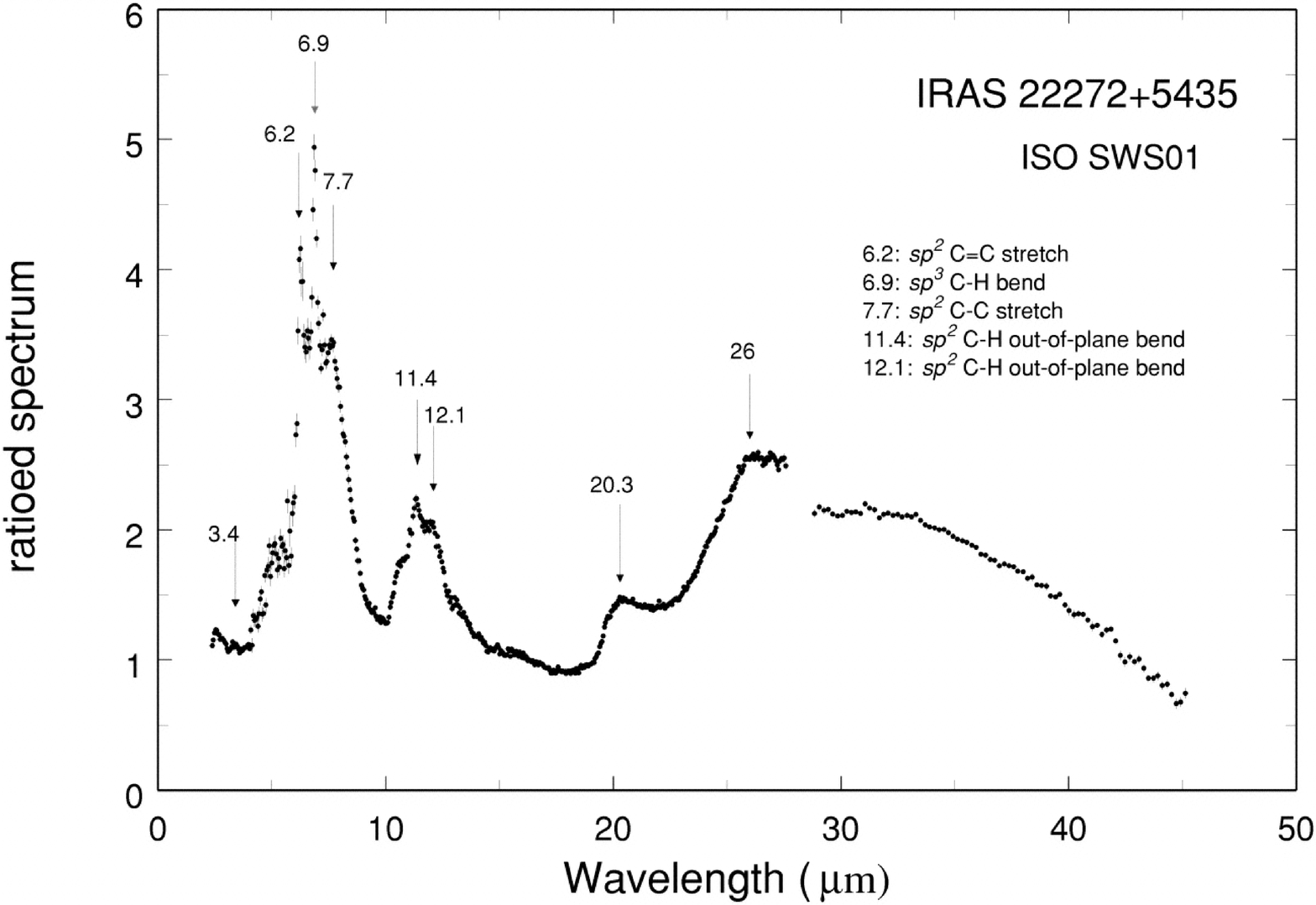}
\end{center}
\caption{\label{label} Left panel: Examples of the several {\it Spitzer} dust
types found in compact (presumably young) Galactic PNe; from top to bottom:
featureless (F); carbon-chemistry aromatic or aliphatic (CCar or CCal);
oxygen-chemistry crystalline or amorphous (OCcr or OCam); and double-chemistry
(DC) (updated from [8]). Right panel: Infrared Space Observatory (ISO) spectrum
of the proto-PN IRAS 22272$+$5435 showing the presence of the still unidentified
21, 26, and 30 $\mu$m features (updated from [14]). The aliphatic discrete
features (e.g., at 3.4 and 6.9 $\mu$m) and the 8 and 12 $\mu$m aliphatic
plateaus together with AIBs (e.g., at 6.2 and 7.7 $\mu$m) are also seen.}
\end{figure}

In general, there is a process of crystallization or aromatization during the
short ($\sim$10$^{2}$-10$^{4}$ years) transition phase AGB$-$PN and the
crystalline (O-rich) and aromatic (C-rich) structures are created during this
short stage (e.g., [10,5,11]). Stars evolving from AGB stars to PNe are thus
great laboratories for astrochemistry, providing us with strong observational
constraints on the gas-phase and solid-state chemical models. The major
challenge is to unveil the formation routes of these complex organic molecules
and inorganic solid-state compounds. Here we present an observational review of
the several molecular (and solid-state) materials that are formed in the
AGB$-$PN phase, focussing on the formation pathways of complex fullerene (and
fullerene-based) molecules as well as on the level of dust processing depending
on metallicity.

\section{Dust composition versus metallicity}

The different {\it Spitzer} dust types observed in PNe are shown in Figure 1
(left panel). These {\it Spitzer} dust types show a strong dependence on
metallicity; see Table 1 in [12] where the {\it Spitzer} dust type statistics
for Galactic bulge (high metallicity) and disk (intermediate metallicity) PNe
are compared with those observed in the lower metallicitiy PNe of the Magellanic
Clouds (MCs). Dust features overimposed to the dust continuum are less frequent
at the lower MC metallicities. Crystalline silicate features (and another O-rich
dust features) are quite rare in the MCs, where carbon chemistry (CC) PNe are
predominant [13]. In addition, double-chemistry (DC) PNe dominate towards the
Galactic bulge [7], while they are not observed in the low-metallcity
environments of the MCs. The mixed-chemistry (DC), C-rich (CC), and O-rich (OC)
PNe are equally distributed in the Galactic disk [8,9].

\subsection{Oxygen-rich dust PNe}

The O-rich dust PNe are less frequent at the lower MC metallicities [13] and the
several {\it Spitzer} O-dust subtypes (amorphous or crystalline) vary with
metallicity (e.g., [9]). In particular, the $\sim$9.7 $\mu$m broad emission band
due to amorphous silicates is more common in the Galactic disk (intermediate
metallicity), while the narrower crystalline silicate emission features (such as
those of olivine and pyroxene at $\sim$23.5, 27.5, and 33.5 $\mu$m) completely
dominate in the Galactic bulge (high metallicity), which otherwise are almost
absent in the MCs. The O-rich dust features evolve from amorphous to crystalline
silicates during the AGB$-$PN phase [10]. Two main scenarios may explain the
silicates crystallization depending on temperature: a) at low temperature in
long-lived circumbinary disks [15]; or b) at high temperature, because of the
strong mass loss, at the tip of the AGB [6].

\subsection{Carbon-rich dust PNe}

A different situation is found for the C-rich dust PNe. They are more common at
the low metallicity of the MCs than at higher metallicity (Galactic disk or
bulge). In the MCs, the C-rich dust PNe generally display small dust grains (or
less processed) [13]; the C-rich aliphatic dust features dominate their {\it
Spitzer} spectra, while the PAH-like emission bands are very rare. In addition,
the C-rich dust PNe dominate at the MC low-metallicity but they are not observed
at the very high metallicity of the Galactic bulge. In particular, broad
emission dust features of aliphatic character (e.g., at $\sim$9$-$13, 15$-$20,
and 25$-$35 $\mu$m; see Section 3) are usually detected in MC PNe, while
Galactic disk C-rich dust PNe are usually compact (presumably young) PNe of
slightly subsolar metallicity [9]. In the case of nearly solar metallicities,
small hydrocarbon molecules (e.g., acetylene) are synthesized in the short
transition phase AGB$-$PN, replacing the broad $\sim$11.5$\mu$m feature due to
SiC (as well as the strong dust continuum emission) seen in the previous AGB
phase. The IR features  rapidly change from aliphatic to PAH-like (aromatic)
during the post-AGB phase and the nebular emission lines (sometimes the broad
$\sim$25$-$35 $\mu$m aliphatic emission feature is still detected) prevail in
the PN stage [10]. The level of dust processing seems to vary with the
metallicity of the environment. For example, the transition from aliphatic
structures (the $\sim$9$-$13, 15$-$20, and 25$-$35 $\mu$m features) to aromatic
ones (the 6.2, 7.7, 8.6, and 11.3 $\mu$m PAH-like features) proceeds more slowly
at the MCs metallicity [13]. Two models for the aromatization process from AGB
stars to PNe have been proposed: a) the PAH precursors are acetylene
(C$_{2}$H$_{2}$) and its radical derivatives [16]; or b) the PAHs are formed as
a consequence of the photochemical processing of the dust grains, which
transform aliphatics to aromatics (e.g., [14,10]). 

\section{Unidentified infrared emission features at 21, 26, and 30 $\mu$m}

The still unidentified set of IR emission features at $\sim$21, 26, and 30
$\mu$m is observed in C-rich sources evolving from AGB stars to PNe (see Figure
1, right panel).

The carrier of the 21 $\mu$m feature, generally observed in post-AGB stars,
should be a fragile solid-state carbon compound (e.g., [17]). Several carriers
such as hydrogenated fullerenes, TiC nanoclusters, nanodiamonds, HAC, and amides
have been proposed in the literature but we still lack a firm identification.

The 30 $\mu$m feature (sometimes with substructure at $\sim$26 $\mu$m), however,
has been linked with magnesium sulfide [18], aliphatic chains [19] or HACs
[20]. The carrier should be quite abundant in the CSE because it is detected
from the AGB to the PN phases, carrying out a significant fraction of the total
energy output. Remarkably, the HAC-like identification, some kind of solid with
a mixed aromatic/aliphatic structure that may also explain the 21 and 26 $\mu$m
features [20], could explain the formation of complex dehydrogenated organics
like C$_{60}$, C$_{70}$, and planar C$_{24}$ (a small piece of a graphene sheet)
in some C-rich PNe (see below).

\begin{figure}
\begin{center}
\includegraphics[width=2.8in]{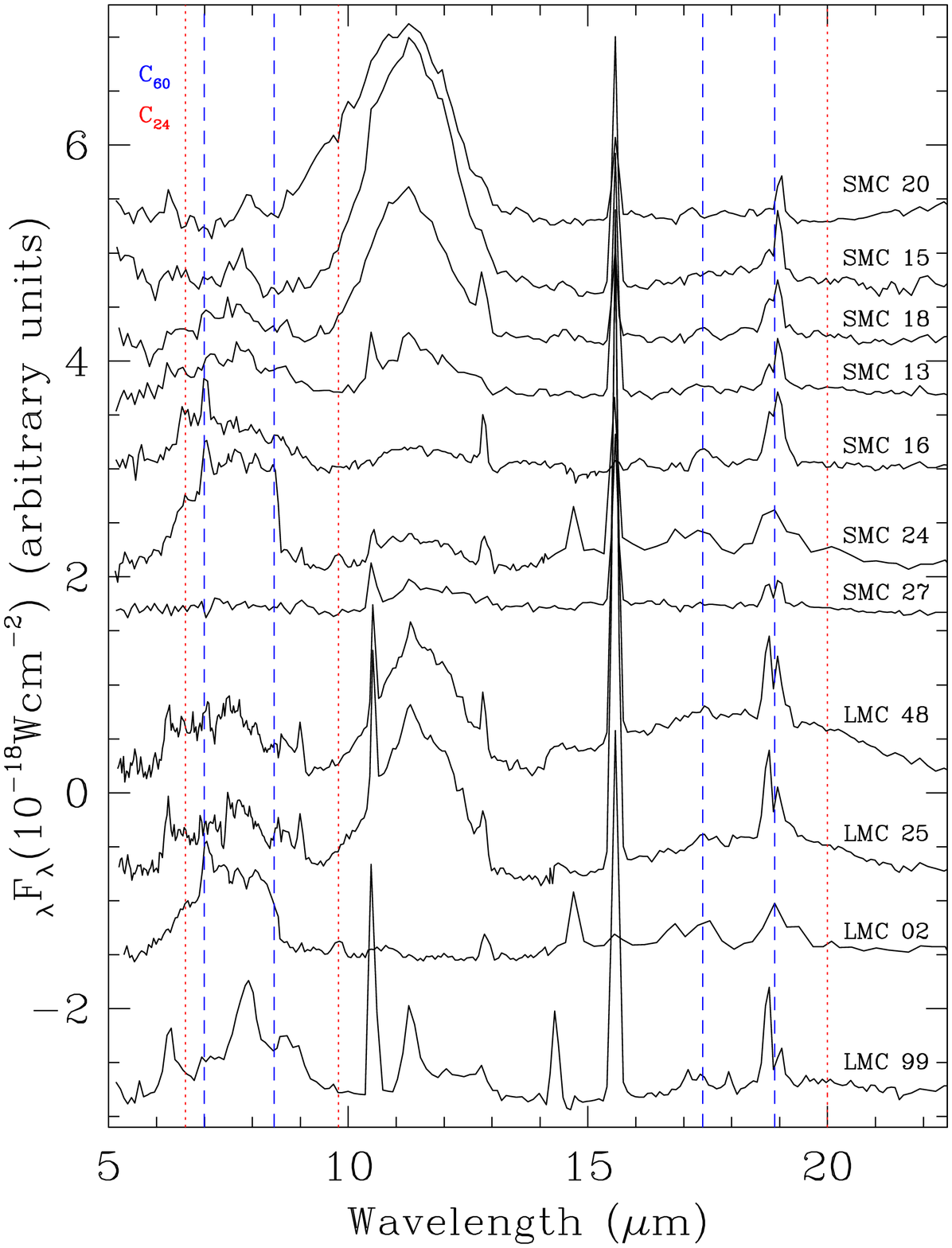}%
\includegraphics[width=3.2in]{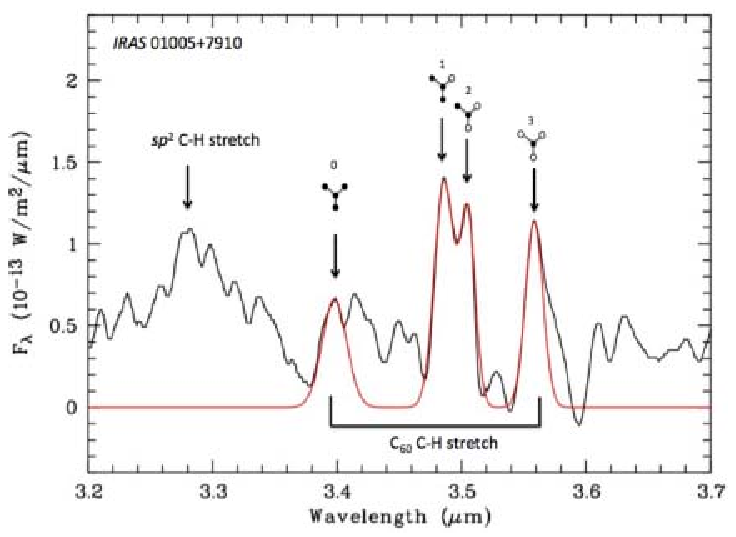}
\end{center}
\caption{\label{label} Left panel: {\it Spitzer} residual spectra of
fullerene-containing PNe in the MCs. The band positions of C$_{60}$ (dashed) and
planar C$_{24}$  (dotted) are marked (updated from [30]). Right panel: ISO
spectrum of the PPN IRAS 01005$+$7910 in the 3-4 $\mu$m spectral region covering
several fullerane features. Note the peaks for the 0, 1, 2, and 3 non-H-bonded
neighboring carbon atoms that may be attributed to fulleranes (updated from
[40]).}
\end{figure}

\section{Detection of complex dehydrogenated molecules}

Kroto et al. [21] discovered the tridimensional C$_{60}$ and C$_{70}$ fullerene
molecules at laboratory and their high stability reinforced the hypothesis that
they should be widespread in the ISM, being important species for the
interstellar/circumstellar chemistry. These complex species may give an
explanation to several phenomena in Astrophysics. For example, they may explain
the diffuse interstellar bands (DIBs) (22; see e.g., [23] for a review on
interstellar/circumstellar DIBs) and the UV bump (e.g., [24]). Fullerenes were
indeed found on Earth and on meteorites and various unsuccessful fullerene
searches (e.g., by using the Infrared Space Observatory) were made to find the
four mid-IR features (at $\sim$7.0, 8.5, 17.4, and 18.9 $\mu$m; see Figure 2,
left panel) of the C$_{60}$ fullerene towards proto-PNe (e.g., [25]) and R
Coronae Borealis (RCB) stars (e.g., [26]). 

Nowadays, C$_{60}$ and C$_{70}$ (first discovered in space by
[27]\footnote{These authors interpreted the detection of C$_{60}$ and C$_{70}$
in the PN Tc 1 as due to the H-poor conditions in the inner core. This
interpretation is in agreement with some laboratory experiments that show that
fullerenes are efficiently produced under the absence of H. However, neither the
central star, nor the inner core and the nebula are H-poor or H-deficient. Thus,
current understanding of stellar astrophysics does not allow for Tc 1 being a
H-poor (late TP) PN (see e.g., [28,29]).}) are known to be efficiently formed in
H-rich circumstellar envelopes only [28,29]. Fullerenes are detected in low-mass
C-rich PNe (both in the Milky Way and in the MCs) with normal H abundances
[28,30] and in only those RCB stars with some H [29]; C$_{60}$ is usually seen
in conjunction with PAH-like features. Indeed, the C$_{60}$ molecule has also
been detected in another astronomical environments such as a PPN [31],
reflection nebulae [32], and young stellar objects [33] and none of them is
H-deficient.

Interestingly, unusual IR emission features at $\sim$6.6, 9.8, and 20 $\mu$m,
attributed to the planar C$_{24}$ molecule (a small piece of a graphene sheet),
have been also detected in MC and Galactic PNe [30,34] and in the ISM [35].
Figure 2 (left panel) shows the planar C$_{24}$ and C$_{60}$ emission features
detected in some PNe. The detection of several complex dehydrogenated organic
molecules like fullerenes and possible planar C$_{24}$ (in conjunction with
aromatic species like PAHs) suggests that other forms of carbon such as
hydrogenated fullerenes (fulleranes like C$_{60}$H$_{36}$ and C$_{60}$H$_{18}$),
fullerene-PAHs adducts, metallofullerenes or even buckyonions and carbon
nanotubes, may be widespread in the Universe. For example: i) the recent studies
of DIBs in C$_{60}$-rich PNe [36,37] suggests the possibility that some DIB
carrier molecules may represent other fullerene-related molecules like
buckyonions; ii) laboratory spectroscopy of adducts between fullerenes and PAHs
(e.g.,C$_{60}$-anthracene/tetracene/pentacene) shows that such fullerene-PAHs
species display the same mid-IR features of isolated C$_{60}$ molecules, being
non distinguishable through mid-IR spectroscopy only [38]; and iii) the recent
non-detections of the most intense IR bands of several fulleranes in two
fullerene PNe [39] together with the (tentative) fulleranes detection in the
proto-PN IRAS 01005$+$7910 [40] (see also Figure 2, right panel) suggest that
fulleranes may be formed in the short transition phase between AGB stars and PNe
but they are quickly destroyed by the UV radiation field from the central star. 

\section{Formation routes of complex dehydrogenated organics}

The formation process of fullerenic and graphenic nanostructures in space is
still unclear to date. Several fullerene formation mechanisms have been
proposed, the most notable ones being: i) the formation in H-poor environments
[41]; ii) high-temperature formation in C-rich environments [42]; iii)
photochemical processing of hydrogenated amorphous carbon grains (HACs; [28]);
and iv) photochemical processing of large PAHs [43]. The first two mechanisms
prevent the formation of hydrogenated species, something that seems to be
difficult to reconcile with the astronomical observations. The alternative
top-down approaches involve the photochemical processing of HACs and large PAHs
as first proposed by [28] and [43], respectively. Both formation scenarios are
based on top-down chemical models towards the most stable C$_{60}$ and C$_{70}$
fullerenes. These top-down fullerene formation scenarios (enumerated below) seem
to be quite different to the fullerene production methods usually employed on
Earth. The terrestrial vaporization or combustion synthesis routes to form
fullerene-like species cannot work in space due to the low gas densities in the
astrophysical environments; e.g., at laboratory, the main formation channel for
C$_{60}$ is the build-up from atomic C, C$_{2}$, small C clusters, and
rings [44].

\subsection{Photochemical processing of HAC-like materials}

Remarkably, fullerene PNe display broad HAC-like dust features at $\sim$9$-$13
and 25$-$35 $\mu$m, suggesting that the most likely explanation for the
simultaneous presence of fullerenes, (possibly) planar C$_{24}$, and PAH-like
emission in the H-rich circumstellar envelopes of PNe is that these molecular
species may be formed from the destruction (e.g., as a consequence of the UV
radiation from the central star) of a carbonaceous compound with a mixture of
aromatic and aliphatic structures - e.g., HACs - , which should a major
constituent in their circumstellar envelopes [28,30,34]. It is to be noted here
that there are other carbonaceous materials with a mixture of aromatic and
aliphatic structures similar to HACs. For example, the mixed aromatic/aliphatic
organic nanoparticles (MAONs) first introduced by Prof. Sun Kwok (e.g.,
[45,46]). The closest natural analogue of such structure is probably kerogen,
which are random arrays of aromatic rings and aliphatic chains with functional
groups made up of H, N, O, and S attached and that may contribute to the
solid-state mid-IR features detected in proto-PNe (e.g., [47]). The coexistence
of a large variety of molecular species such as HACs, PAH clusters, fullerenes,
and small dehydrogenated carbon clusters (possibly planar C$_{24}$ or graphene
precursors) in PNe with fullerenes strongly supports the few laboratory
experiments carried out by Scott and colleagues in the nineties, which showed
that the decomposition of HACs is sequential with small dehydrogenated PAH
molecules being released first, followed by fullerenes and large PAH clusters
[48]. This fullerene formation scenario is also suggested by the strong - and
unique - spectral variations (in a timescale of just a few years) seen in the IR
spectrum of the fullerene-containing RCB star V854 Cen, which indicate that a
significant fraction of the dust grains in the envelope have evolved from HACs
to complex species such as PAHs and fullerene-like molecules [29]. The recent
works by [49,50] also suggest that fullerene formation in PNe likely starts from
HAC processing. Perhaps the most novel idea is introduced by [50] who propose
that UV-photolysis produces structural changes in the HAC grains, forming 3D
hollow structures or arophatic clusters (aromatic clusters linked by aliphatic
bridging groups). These arophatic clusters end up with a cage-like, cup-like or
tube-like aspect, being curled-up or folded-over graphene sheets. Further
UV-induced dehydrogenation of these arophatic clusters may introduce pentagonal
rings, permitting the curvature of the structure into large fullerenic cages
that can shrink down to the most stable C$_{60}$ and C$_{70}$ fullerene
configurations.

\subsection{Photochemical processing of large PAHs}

This fullerene formation route has been proposed to explain the formation of
fullerenes in the ISM [43]. These authors propose an alternative top-down model
where large PAHs (with $\sim$70 C atoms) are converted into graphene, and
subsequently fullerenes, under the action of UV photons from massive stars. A
schematic view of this alternative top-down fullerene formation model is shown
in Figure 3 of [43]. Upon UV irradiation, graphene formation takes place through
PAH photolysis. The expected dehydrogenation, fragmentation, and isomerization
can give rise to a rich organic chemistry, forming fullerenes, small cages,
rings, and chains. Carbon loss followed by pentagonal defects formation that
induces curvature of the graphene sheet may form fullerenes by migration of the
pentagons. Remarkably, planar molecules such as C$_{20}$ - C$_{30}$ are expected
to be formed by further graphene fragmentation. In the reflection nebula NGC
7023, emission at 6.6 $\mu$m (attributed to possible planar C$_{24}$ molecules)
is seen near the ionizing star [35], suggesting that the carrier is a
photo-product of PAHs. In this context, the possible detection of the very
stable planar C$_{24}$ molecule in the ISM is also consistent with the idea of
fullerene production from the photochemical processing of PAHs via graphene
formation.

\section{Complex dehydrogenated organics versus metallicity}

Interestingly, the detection rate of complex dehydrogenated fullerene molecules
in C-rich PNe decreases with increasing metallicity. In our own Galaxy, only
$\sim$5\% of the PNe display the fullerene features, which otherwise are clearly
detected in $\sim$20\% and $\sim$44\% of the PNe observed in the LMC and SMC,
respectively [34]. This interesting finding suggests a higher presence of small
dust grains (or a more limited dust processing) at the lower MCs metallicity.
More recently, [51] demonstrate that all Galactic fullerene-containing PNe are
low-mass objects of sub-solar metallicity, showing that low metallicity
environments are more favourable to fullerene production and detection. 

On the other hand, the still unidentified 21 $\mu$m feature is less frequent in
the Galaxy than in the MCs [52] and the 21 $\mu$m carrier could be
related with the formation of fullerenes. In particular, an anti-correlation
between the 30 $\mu$m and other UIR features is seen among the 21 $\mu$m sources
in the MCs [52] and such anti-correlation could result from
the photochemical processing of HAC-like dust grains into fullerenes and PAHs.
Under the HACs hypothesis for fullerene formation, the 21 $\mu$m feature should
be related with the formation of fullerenes with the carrier being a fragile
intermediate product from the decomposition of HAC or a similar material with
mixed aromatic/aliphatic structures. However, [53], although
based on a smaller sample of Galactic post-AGB stars, reported that the 30
$\mu$m feature and the other UIRs are not anti-correlated in Galactic 21
$\mu$m objects. Interestingly, the Galactic 21 $\mu$m emitters are
spectroscopically different to the MC ones; i.e., the 21 $\mu$m proto-PNe in the
MCs show more (typical) PAH-like aromatic features.

\section{Concluding remarks}

The IR dust characteristics (and evolution) observed from the AGB to the PN
phases in different metallicity environments (Galactic disk, bulge, and MCs)
agree well with the predictions of the AGB nucleosynthesis models (i.e., the
expected effects of the third dredge-up and HBB processes). The coexistence of
aromatic PAH-like species and complex dehydrogenated organic molecules such as
fullerenes and planar C$_{24}$ (a small piece of a graphene sheet) suggests a
top-down model for the formation of fullerenes in space. Low metallicity
environments like those of the MCs favour fullerene formation and detection as
well as a more limited dust processing (or the general presence of small dust
grains). Although more observational and laboratory efforts are needed, it
seems clear that a diverse family of fullerene-related molecules (e.g.,
fulleranes, metallofullerenes, fullerene-PAH adducts, buckyonions, etc.) is very
likely to be present in space.

\ack
D.A.G.H. was funded by the Ram\'on y Cajal fellowship number RYC$-$2013$-$14182.
D.A.G.H and A. M. acknowledge support provided by the Spanish Ministry of Economy and
Competitiveness (MINECO) under grant AYA$-$2014$-$58082-P.

\smallskip

\end{document}